\documentstyle[12pt]{article}
\begin{document}

\hspace*{60ex} OCU-PHYS-245 

\hspace*{60ex} April 2006 

\hspace*{60ex} July 2006 (Revised) 

\begin{center}
{\Large {\bf 
Renormalization in Coulomb-gauge QCD within the Lagrangian formalism 
} 
} \\

\hspace*{3ex} 

\hspace*{3ex}

{\large 
{\sc A. Ni\'{e}gawa}

{\normalsize\em Graduate School of Science, Osaka City University} 
\\ 
{\normalsize\em Sumiyoshi-ku, Osaka 558-8585, Japan} } \\

\hspace*{2ex}

\hspace*{2ex}

\hspace*{2ex}

{\large {\bf Abstract}} \\ 
\end{center}
We study renormalization of Coulomb-gauge QCD within the Lagrangian, 
second-order, formalism. We derive a Ward identity and the 
Zinn-Justin equation, and, with the help of the latter, we give a 
proof of algebraic renormalizability of the theory. Through 
diagrammatic analysis, we show that, in the strict Coulomb gauge, 
$g^2 D^{00}$ is invariant under renormalization. ($D^{00}$ is 
the time-time component of the gluon propagator.)

\newpage
%%%%%%%%%%%%%%%%%%%%%%%%%%%%%%%%%%%%%%%%%%%%%%%%%%%%%%
%%%%% SEC %%%%%%%%%%%%%%%%%%%%%%%%%%%%%%%%%%%%%%%%%%%%
%%%%%%%%%%%%%%%%%%%%%%%%%%%%%%%%%%%%%%
\section{Introduction}
In nonabelian gauge theories, among a variety of gauge choices, the 
Coulomb gauge is one of the most important ones. The theories with 
this gauge are described in terms of physical fields, so that the 
unitarity is manifest. Within the Hamiltonian, first-order, 
formalism, formal or algebraic renormalizability of the Coulomb 
gauge has been studied in \cite{Z1}. Since then, introducing an 
interpolating gauge, which interpolate between a covariant gauge and 
the Coulomb gauge, Baulieu and Zwanziger have proved algebraic 
renormalizability of the theory \cite{Z2}. In taking the 
Coulomb-gauge limit, a phase-spaces representation is used there. 
Despite its importance, the proof of renormalizability of the 
Coulomb gauge {\em within the Lagrangian, second-order, formalism 
per se} is still lacking, to which the present paper is devoted. 

In this paper, we are concerned about Coulomb gauge QCD, whose 
Lagrangian density is obtained by adding the gauge fixing term, 
${\cal L}_{\mbox{\scriptsize{G.F.}}} = - (\partial_i A_a^i) 
(\partial_j A_a^j) / 0^+$, to the QCD Lagrangian density. 
In the Coulomb gauge, there is an inherent problem of appearance of 
\lq energy divergences', which are characteristic of an 
instantaneous closed ghost- and $A_0$-loops: 
%%%%%%%%%%%%%%%%%%%%%%%%%%%%%%%%%%%%%%%%%%%%%%%%%%%%%%%%%%%
\begin{equation}
\int \frac{d p_0}{2 \pi} \, F ({\bf p}, ... ) \, , 
\label{Einf} 
\end{equation}
%%%%%%%%%%%%%%%%%%%%%%%%%%%%%%%%%%%%%%%%%%%%%%%%%%%%%%%
where $F$ is independent of $p_0$, the temporal component of the 
four-vector $P^\mu = ( p_0, {\bf p})$ and \lq\lq $...$'' indicates a 
set of external momenta. There also appear ill-defined integrals of 
the forms, 
%%%%%%%%%%%%%%%%%%%%%%%%%%%%%%%%%%%%%%%%%%%%%%%%%%%%%%%%%%%
\begin{eqnarray}
&& \int \frac{d p_0}{2 \pi} \, \frac{p_0}{p_0^2 - p^2 + i 0^+} \, 
G ({\bf p}, ... ) \, , 
\label{ill1} \\ 
&& \int \frac{d p_0}{2 \pi} \frac{d q_0}{2 \pi} \, 
\frac{p_0}{p_0^2 - p^2 + i 0^+} \, 
\frac{q_0}{q_0^2 - q^2 + i 0^+} \, H ({\bf p}, {\bf q}, ... ) \, , 
\label{ill2} 
\end{eqnarray}
%%%%%%%%%%%%%%%%%%%%%%%%%%%%%%%
where $G$ ($H$) is independent of $p_0$ ($p_0$ and $q_0$). 

A numerous work has been devoted to the energy divergence problem 
and, by now, the following results are established. 
\begin{description} 
\item{1)} In the Hamiltonian, phase-space, first order, formalism, 
energy divergences like (\ref{Einf}) do not appear \cite{doust} 
{\em in the first place}.  
\item{2)} Using the correspondence formula which equates amplitudes 
in a covariant gauge to those in a gauge without ghosts, Cheng and 
Tsai \cite{CT} \lq\lq indirectly'' showed that, when all relevant 
contributions are added, cancellation occurs between the energy 
divergences (\ref{Einf}). 
\item{3)} With the help of an interpolating gauge, which 
interpolates between a covariant gauge and the Coulomb gauge, in the 
phase-space formalism, it has been shown by Baulieu and Zwanziger 
\cite{Z2} that the cancellation occurs between different 
contributions which turn out to energy-divergent ones in the 
Coulomb-gauge limit (see, also, \cite{doust}). 
\item{4)} Ill-defined integrals like (\ref{ill1}) can be set equal 
to zero \cite{CT2} and another type of ill-defined integrals, Eq. 
(\ref{ill2}), are connected \cite{doust,CT2} with the so-called 
$V_1 + V_2$ terms of Christ and Lee \cite{CL}, which arise through 
correct treatment of operator ordering in the Hamiltonian. 
\item{5)} It has been shown in \cite{AT} that the cancellation of 
energy divergences and renormalizability is compatible in an 
example in which quark-loop subgraphs are inserted into the 
second-order gluon self-energy graphs: As mentioned in 1) above, 
in the phase-space formalism, two integrals over the internal 
energies converge. However, in relation to renormalization, 
energy-divergences re-appear. Thanks to the Ward identity, these 
energy-divergent contributions cancel out. 
\end{description} 
Therefore, when perturbative computations in the present Lagrangian 
formalism are properly handled, cancellation should occur between 
the energy divergences. We are not concerned, in this paper, with 
the energy-divergence issue anymore. 

As stated above, we are are interested in the Coulomb gauge QCD. 
Nevertheless, we proceed, as far as possible, with more general 
gauge choice, ${\cal L}_{\mbox{\scriptsize{G.F.}}} = - (\partial_i 
A_a^i)^2 / (2 \alpha)$ with arbitrary $\alpha$, 
which is usually called the Coulomb gauge. The \lq\lq genuine'' 
Coulomb gauge, which is obtained by taking the limit $\alpha \to 
0^+$, is called the strict Coulomb gauge. Incidentally, in the 
case of $\alpha \neq 0$ Coulomb gauge, when compared to the strict 
Coulomb gauge, much worse energy divergences arise (cf. Eq. 
(\ref{cir}) below), so that, at the present stage, such a gauge is 
useless for practical perturbative calculations. 

In \S2, we present a Ward identity and the Zinn-Justin equation, 
which are derived in Appendix A. In \S3, on the basis of the 
Zinn-Justin equation, we construct renormalization counterterms in a 
recursive way. In \S4, we prove an algebraic renormalizability 
of the theory and obtain the identities between the renormalization 
constants. With the aid of the Ward identity, we obtain an 
additional identity $\tilde{Z}_1 = Z_{12} / Z_{31}$ (Eq. 
(\ref{hiya})). We then show that, in the strict Coulomb gauge, $g^2 
D^{00}$ ($D^{00}$ is the time-time component of the gluon 
propagator) is invariant under renormalization. In \cite{Z1,Z2}, 
this proposition is proved using the Ward identity. In contrast, in 
the present Lagrangian formalism, use of the Zinn-Justin equation is 
sufficient for proving this. For illustrative purpose, the forms of 
some one-loop renormalization constants in the strict Coulomb gauge 
are also displayed. \S5 is devoted to summary. Appendix A briefly 
describes a derivation of the Ward identity and the Zinn-Justin 
equation. In Appendix B, using the identity that is dereived from 
the Zinn-Justin equation, we obtain, in the strict Coulomb gauge, an 
additional identities $\tilde{Z}_1 = D' = 1$, Eq. (\ref{BB}). In 
Appendix C, we briefly describe a derivation of some one-loop 
renormalization constants in the strict Coulomb gauge. Some formal 
diagrammatic analyses are summarized in Appendix D. In particular, 
we show that, in the strict Coulomb gauge, $\bar{\eta} A 
\eta$-vertex (cf. Eq. (\ref{lag})) is not a renormalization part, so 
that $\tilde{Z}_1 = 1$ in the minimal subtraction scheme in 
dimensional regularization. 
%%%%%%%%%%%%%%%%%%%%%%%%%%%%%%%%%%%%%%%%%%%%%%%%
%%%%%%%% SEC %%%%%%%%%%%%%%%%%%%%%%%%%%%%%%%%%%%%%%%%%
%%%%%%%%%%%%%%%%%%%%%%%%%%%%%%%%%%%%%%%%%%
\section{Ward identity and Zinn-Justin equation}
As the content of this section is standard, we describe briefly. 
Greek indices $\mu$, $\nu$, ... run over $0, 1, 2, 3$, while Latin 
indices $i$, $j$ run over $1, 2, 3$. We use $P^\mu$ for denoting a 
four vector $P^\mu = (p^0, {\bf p})$ and $p^j$ for denoting a three 
vector. 

The effective Lagrangian density of Coulomb-gauge QCD with one quark 
flavor (generalization to the case of several quarks is 
straightforward) reads 
%%%%%%%%%%%%%%%%%%%%%%%%%%
\begin{eqnarray}
{\cal L}_{\mbox{\scriptsize{eff}}} &=& 
\tilde{\cal L}_{\mbox{\scriptsize{eff}}} - \frac{1}{2 \alpha} 
{\cal F}_a {\cal F}_a \;\;\;\;\;\; \left( {\cal F}_a [A; x] = 
\partial_i A^i_a (x) \right) \, , \nonumber \\ 
\tilde{\cal L}_{\mbox{\scriptsize{eff}}} &=& - \frac{1}{4} 
F^{\mu \nu}_a F_{a \mu \nu} + \bar{\psi} \left( i 
\partial\kern-0.045em\raise0.3ex\llap{/}\kern0.25em\relax - m - g 
t_a {A\kern-0.1em\raise0.3ex\llap{/}\kern0.35em\relax}_a \right) 
\psi + \partial_i \bar{\eta}_a D^i_{a b} (x) \eta_b \, , 
\label{lag} 
\end{eqnarray}
%%%%%%%%%%%%%%%%%%%%%%%%%%%%%
where $F^{\mu \nu}_a = \partial^\mu A^\nu_a - \partial^\nu A^\mu_a - 
g f_{a b c} A^\mu_b A^\nu_c$, $t_a = \lambda_a / 2$, and $D_{ab}^\mu 
(x) = \delta_{ab} \partial^\mu + g f_{abc} A^\mu_c (x)$. 
Generalization to other nonabelian gauge theories is straightforward. 
In the Lagrangian formalism adopted here, the fields propagators can 
be extracted from the bilinear (with respect to the fields) terms of 
${\cal L}_{\mbox{\scriptsize{eff}}}$ in Eq. (\ref{lag}). For the 
purpose of later use, among the propagators, we only display the 
forms of the gluon propagator $\Delta^{\mu \nu} (Q)$ (see, also, 
\cite{bra,and1}) and FP-ghost propagator $\tilde{\Delta} (Q)$, 
together with $\bar{\eta}_a (Q) A_b^i \eta_c$-vertex factor 
${\cal V}_{abc}^i (Q)$: 
%%%%%%%%%%%%%%%%%%%%%%%%%%
\begin{eqnarray}
\Delta^{\mu \nu} (Q) &=& g^{\mu}_{\; i} g^{\nu}_{\; j} 
\frac{\delta^{i j} - q^i q^j / q^2}{Q^2 + i 0^+} + \frac{n^\mu 
n^\nu}{q^2} - \alpha \frac{Q^\mu Q^\nu}{q^4} \, , 
\label{cir} \\ 
\tilde{\Delta} (Q) &=& - \frac{1}{q^2} \, , 
\label{tri} \\ 
{\cal V}_{abc}^i (Q) &=& g f_{a b c} q^i \, . 
\label{star}
\end{eqnarray}
%%%%%%%%%%%%%%%%%%%%%%%%%
Propagators are diagonal in color space, so that the color indices 
are suppressed. 

The gauge-field part and the quark part of 
$\tilde{\cal L}_{\mbox{\scriptsize{eff}}}$ is invariant under the 
infinitesimal gauge transformation: $A_a^\mu \to A_a^\mu + 
D^\mu_{ab} (x) \epsilon_b (x) $, $\psi \to \psi - i g \epsilon_a 
t_a \psi$, and $\bar{\psi} \to \bar{\psi} + i g \bar{\psi} 
\epsilon_a t_a$. ${\cal L}_{\mbox{\scriptsize{eff}}}$ is invariant 
under the BRST transformation of the fields 
%%%%%%%%%%%%%%%%%%%%%%%%%%
\begin{eqnarray}
\delta A^\mu_a (x) &=& D^\mu_{ab} (x) \eta_b \zeta \equiv sA^\mu_a 
\zeta \, , 
\label{BRS2} \\ 
\delta \eta_a (x) &=& - \frac{g}{2} f_{abc} \eta_b (x) \eta_c (x) 
\zeta \equiv s\eta_a \zeta \, , 
\label{BRS3} \\ 
\delta \psi & = & i g \eta_a t_a \psi \zeta \equiv s\psi \zeta \, , 
\;\;\;\; \delta \bar{\psi} = i g \bar{\psi} \eta_a t_a 
\zeta \equiv s\bar{\psi} \zeta \, , \label{BRS1} \\ 
\delta \bar{\eta}_a (x) &=& \frac{1}{\alpha} {\cal F}_a [A; x] \zeta 
\equiv s\bar{\eta}_a \zeta \, . 
\label{BRS4} 
\end{eqnarray}
%%%%%%%%%%%%%%%%%%%%%%%%%%%%%
Here $\zeta$ is an $x$-independent infinitesimal Grassmann number 
with the same ghost number as $\bar{\eta}_a$. 
${\cal L}_{\mbox{\scriptsize{eff}}}$ is not invariant under the 
Lorentz transformation but is invariant under the spatial rotation. 
Then, we treat the spatial component $A_a^i$  and the temporal 
component $A_a^0$ of $A_a^\mu$ separately. We let $\chi^n$ run over 
the fields, $A_a^i$, $A_a^0$, $\eta_a$, $\psi$, and $\bar{\psi}$, 
and introduce a set of c-number external sources $K_n (x)$ that 
couple to $s \chi^n [\chi; x]$, $K_n = (K_a^i, K_a^0, K_{\eta_a}, 
K_\psi, K_{\bar{\psi}})$. 

Quantum effective action $\Gamma$ is defined by the following 
implicit functional integro-differential equation: 
%%%%%%%%%%%%%%%%%%%%%%%%%%
\begin{eqnarray}
e^{i \Gamma [\chi, \bar{\eta}, K]} & = & \int {\cal D} (\chi', 
\bar{\eta}') \exp \left[ i \int d^4 y \left( 
{\cal L}_{\mbox{\scriptsize{eff}}} (\chi', \bar{\eta}') 
+ K_n s \chi^{' n} 
\right. \right. \, \nonumber \\ 
&& \left. \left. - \frac{\delta_R \Gamma}{\delta \chi^n} 
\left( \chi^{' n} - \chi^n \right) - \frac{\delta_R \Gamma}{\delta 
\bar{\eta}_a} \left( \bar{\eta}_a' - \bar{\eta}_a \right) \right) 
\right] \, , 
\label{chu}
\end{eqnarray}
%%%%%%%%%%%%%%%%%%%%%%%%%%%%%
where \lq $R$' denotes right differentiation. We show in Appendix A 
that 
%%%%%%%%%%%%%%%%%%%%%%%%%%
\begin{equation}
\Gamma [\chi, \bar{\eta}, K] \mbox{ depends on } \bar{\eta} 
\mbox{ only through } K_a^j + \partial^j \bar{\eta}_a \, . 
\label{imp}
\end{equation}
%%%%%%%%%%%%%%%%%%%%%%%%%%%%%
We introduce $\tilde{\Gamma}$ through 
%%%%%%%%%%%%%%%%%%%%%%%%%%
\begin{equation}
\tilde{\Gamma} = \Gamma + \frac{1}{2 \alpha} \int d^4 y \, 
{\cal F}^2 [A; y] \, . 
\label{henkou}
\end{equation}
%%%%%%%%%%%%%%%%%%%%%%%%%%%%%
The Ward identity, which is derived in Appendix A, reads 
%%%%%%%%%%%%%%%%%%%%%%%%%%
\begin{equation}
\int d^3 x \left[ \frac{\delta_R \tilde{\Gamma}}{\delta \chi^n} 
\frac{\delta_L \tilde{\Gamma}}{\delta K_n} \right] (x) = 
\partial_0 \int d^3 x \left[ \eta_a \frac{\delta_R 
\tilde{\Gamma}}{\delta A_a^0} + K_a^0 \frac{\delta_L 
\tilde{\Gamma}}{\delta K_{\eta_a}} \right] (x) \, , 
\label{chu4te}
\end{equation}
%%%%%%%%%%%%%%%%%%%%%%%%%%%%%
where \lq $L$' denotes left differentiation. Integration over $x_0$ 
yields the Zinn-Justin equation, 
%%%%%%%%%%%%%%%%%%%%%%%%%%
\begin{equation}
0 = \int d^4 x \left[ \frac{\delta_R \tilde{\Gamma}}{\delta \chi^n} 
\frac{\delta_L \tilde{\Gamma}}{\delta K_n} \right] (x) \equiv \left( 
\tilde{\Gamma}, \; \tilde{\Gamma} \right) \, . 
\label{ST6}
\end{equation}
%%%%%%%%%%%%%%%%%%%%%%%%%%%%%
The leading term (in the loop expansion) of $\tilde{\Gamma}$ is 
%%%%%%%%%%%%%%%%%%%%%%%%%%
\begin{equation}
\tilde{\Gamma}_0 [\chi, \bar{\eta}, K] = \int d^4 x \left( 
\tilde{\cal L}_{\mbox{\scriptsize{eff}}} [\chi, \bar{\eta},; x] + 
K_n s\chi^n [\chi; x] \right) \, ,  
\label{zero} 
\end{equation}
%%%%%%%%%%%%%%%%%%%%%%%%%%%%%
which is invariant under the transformation (\ref{BRS2}) - 
(\ref{BRS1}). It should be noted that the fields $\chi^n$ and 
$\bar{\eta}$ in $\tilde{\Gamma}_0$ are the renormalized ones. 
%%%%%%%%%%%%%%%%%%%%%%%%%%%%%%%%%%%%%%%%%%%%%%%%%%%%%%%
%%%%%%% SEC %%%%%%%%%%%%%%%%%%%%%%%%%%%%%%%%%%%%%%%%%%%
%%%%%%%%%%%%%%%%%%%%%%%%%%%%%%%
\section{Recursive construction of counterterms} 
In this section, we construct the renormalization counterterms that 
preserve the symmetry condition (\ref{ST6}). We follow the procedure 
in standard text books \cite{IZ,wein}, so that we briefly describe. 

Let us use dimensional regularization by continuing spacetime 
dimensions from $4$ to $d$. We employ the loop expansion for 
$\tilde{\Gamma}$, $\tilde{\Gamma} = \sum_{N = 0}^\infty 
\tilde{\Gamma}_N$ and decompose $\tilde{\Gamma}_N$ as 
$\tilde{\Gamma}_N = \tilde{\Gamma}_{R N} + 
\tilde{\Gamma}_N^{(\infty)}$, where $\tilde{\Gamma}_N^{(\infty)}$ is 
the ultra-violet (UV) divergent contribution, i.e., it diverges in 
the limit $d \to 4$. We adopt the minimal subtraction scheme. 

The symmetry condition (\ref{ST6}) leads to $\sum_{N' = 0}^N$ 
$\left( \tilde{\Gamma}_{N'}, \right.$ $\left. \; 
\tilde{\Gamma}_{N - N'} \right)$ $= 0$ $(N$ $=$ 0, \,$ $1, \,$ 
$2,$ ... )$. We proceed in a recursive way. 
We assume that, for all $M \leq N - 1$, all UV-divergent 
contributions from $M$-loop diagrams have been cancelled by 
counterterms $\tilde{\Gamma}_M^{(\infty)}$ $(M$ $=$ $1, $ $2,$ 
$ ... ,$ $N - 1)$. (This recursive procedure is justified 
{\em a posteriori} in \S4.) Then UV infinities can appear only 
in the $N' = 0$ and $N' = N$ terms, and the infinite part of the 
above condition reads $\left( \tilde{\Gamma}_0, \right.$ $\left. 
\tilde{\Gamma}_N^{(\infty)} \right)$ $+ \left( \Gamma_N^{(\infty)}, 
\right.$ $\left. \; \tilde{\Gamma}_0 \right)$ $= 0$. Here, 
$\tilde{\Gamma}_0$ is as in Eq. (\ref{zero}). We write 
$\tilde{\Gamma}^{(\infty)}_N$ as 
%%%%%%%%%%%%%%%%%%%%%%%%%%
\begin{equation}
\tilde{\Gamma}^{(\infty)}_N [\chi, \bar{\eta}, K] = \int d^4 x 
\left( \tilde{l}_N [\chi; x] + K_n {\cal D}^n_N [\chi; x] \right) 
\, , 
\label{enu} 
\end{equation}
%%%%%%%%%%%%%%%%%%%%%%%%%%%%%
where $\tilde{l}_N$ and ${\cal D}^n_N$ are local functions of 
$\chi^n$ and $\bar{\eta}$ and their derivatives \cite{wein}. 
For the time being, we drop the suffix $N$. Substitution of Eqs. 
(\ref{zero}) and (\ref{enu}) gives 
%%%%%%%%%%%%%%%%%%%%%%%%%%
\begin{eqnarray}
&& \int d^4 x \left( \frac{\partial_R 
\tilde{\cal L}_{\mbox{\scriptsize{eff}}} [\chi, \bar{\eta}; 
x]}{\partial \chi^n (x)} {\cal D}^n [\chi; x] + 
\frac{\partial_R \tilde{l} [\chi, \bar{\eta}; x]}{\partial 
\chi^n (x)} s\chi^n [\chi; x] \right) = 0 \, , 
\label{daiiti} \\ 
&& \int d^4 x \, \left( \frac{\partial_R \, s\chi^m 
[\chi; x]}{\partial \chi^n (x)} {\cal D}^n [\chi; x] + 
\frac{\partial_R {\cal D}^m [\chi; x]}{\partial \chi^n (x)} 
s\chi^n [\chi; x] \right) = 0 \, . 
\label{daini}
\end{eqnarray}
%%%%%%%%%%%%%%%%%%%%%%%%%%%%%

From here on, we follow the procedure in Sec. 17 of \cite{wein}. 
We introduce 
%%%%%%%%%%%%%%%%%%%%%%%%%%
\begin{eqnarray}
\tilde{\Gamma}^{(\epsilon)} [\chi, \bar{\eta}] &=& \int d^4 x \left( 
\tilde{\cal L}_{\mbox{\scriptsize{eff}}} [\chi, \bar{\eta}; x] + 
\epsilon \tilde{l} [\chi, \bar{\eta}; x] \right) \nonumber \\ 
& \equiv & \int d^4 x \, \tilde{\cal L}^{(\epsilon)} [\chi, 
\bar{\eta}; x] \, , 
\label{cat} \\ 
\Delta^{(\epsilon) n} [\chi; x] &=& s\chi^n [\chi; x] + \epsilon 
{\cal D}^n [\chi; x] \, , 
\label{4.5} 
\end{eqnarray}
%%%%%%%%%%%%%%%%%%%%%%%%%%%%%
with $\epsilon$ infinitesimal for technical reason. 
Then, Eqs. (\ref{daiiti}) says that $\tilde{\Gamma}^{(\epsilon)}$ is 
invariant under the transformation, 
%%%%%%%%%%%%%%%%%%%%%%%%%%
\begin{equation}
\chi^n (x) \to \chi^n (x) 
+ \Delta^{(\epsilon) n} [\chi; x] \zeta \, , 
\label{henkan}
\end{equation}
%%%%%%%%%%%%%%%%%%%%%%%%%%%%%
while Eq. (\ref{daini}) tells us that this transformation is 
{\em nilpotent}. 

The most general form of the transformation (\ref{henkan}) is 
%%%%%%%%%%%%%%%%%%%%%%%%%%
\begin{eqnarray*}
A_a^i & \to & A_a^i + \left[ B_{ab}^{(\epsilon)} \partial^i \eta_b 
- g D_{abc}^{(\epsilon)} A_b^i \eta_c \right] \zeta \, , \\ 
A_a^0 & \to & A_a^0 + \left[ B_{ab}^{(\epsilon) '} \partial^0 \eta_b 
- g D_{abc}^{(\epsilon) '} A_b^0 \eta_c \right] \zeta \, , \\ 
\eta_a & \to & \eta_a - \frac{g}{2} E_{abc}^{(\epsilon)} \eta_b 
\eta_c \zeta \, , \\ 
\psi & \to & \psi + i g \eta_a T_a^{(\epsilon)} \psi \zeta \, , 
\end{eqnarray*}
%%%%%%%%%%%%%%%%%%%%%%%%%%%%%
where $B_{ab}^{(\epsilon)}$, $B_{ab}^{(\epsilon) '}$, 
$D_{abc}^{(\epsilon)}$, $D_{abc}^{(\epsilon) '}$, and 
$E_{abc}^{(\epsilon)}$ $(= - E_{acb}^{(\epsilon)})$ are constants, 
and $T_a^{(\epsilon)}$ is some matrix acting on the quark field. 
Here $D^{(\epsilon)}_{abc} = (D^{(\epsilon)}_{abc})_N = f_{abc} 
+ \epsilon (D_{abc})_N$, $T_a^{(\epsilon)} = (T_a^{(\epsilon)})_N = 
t_a + \epsilon (T_a)_N$, etc. Even if we introduce different 
$T_a^{(\epsilon)}$'s for different components of $\psi$, we will 
have the same result as the one obtained below. Imposing the 
condition of nilpotence, we obtain $E_{abc}^{(\epsilon)} = 
D_{abc}^{(\epsilon)} = D_{abc}^{(\epsilon) '} = C^{(\epsilon)} 
f_{abc}$, $B_{ab}^{(\epsilon)} = (C D)^{(\epsilon)} \delta_{ab}$, 
$B_{ab}^{(\epsilon) '} = (C D')^{(\epsilon)} \delta_{ab}$, and 
$T_a^{(\epsilon)} = C^{(\epsilon)} t_a$, where $C^{(\epsilon)}$, 
$(CD)^{(\epsilon)}$, and $(CD')^{(\epsilon)}$ are some constants: 
$C^{(\epsilon)} = C^{(\epsilon)}_N = 1 + \epsilon C_N$, 
$(CD)^{(\epsilon)} = (CD)^{(\epsilon)}_N = 1 + \epsilon \sum_{N' = 
0}^N C_{N'} D_{N - N'}$ ($C_0 = D_0 = 1$), etc. ($C_M$ and $D_M$ $(M 
< N)$ have already been determined at lower-order stages.) Thus, we 
have, with obvious notation, 
%%%%%%%%%%%%%%%%%%%%%%%%%%
\begin{eqnarray}
A_a^i & \to & A_a^i + \left[ C D \left( \partial^i \eta_a - 
\tilde{g} f_{abc} A_b^i \eta_c \right) \zeta \right]^{(\epsilon)} 
\, , 
\label{eei1} \\ 
A_a^0 & \to & A_a^0 + \left[ C D \left\{ (D' / D) \partial^0 \eta_a 
- \tilde{g} f_{abc} A_b^0 \eta_c \right\} \zeta \right]^{(\epsilon)} 
\, , 
\label{ee01} \\ 
\eta_a & \to & \eta_a + \left[ C D \left( - \frac{1}{2} \tilde{g} 
f_{abc} \eta_b \eta_c \zeta \right) \right]^{(\epsilon)} \, , 
\label{eta1} \\ 
\psi & \to & \psi + \left[ C D \left( i \tilde{g} \eta_a t_a \psi 
\zeta \right) \right]^{(\epsilon)} \, , 
\label{pusai1} 
\end{eqnarray}
%%%%%%%%%%%%%%%%%%%%%%%%%%%%%
where $\tilde{g} \equiv g / D$. Introducing $\tilde{A}_a^0 \equiv 
\left( D / D' \right) A_a^0$, one can rewrite Eq. (\ref{ee01}) as 
%%%%%%%%%%%%%%%%%%%%%%%%%%%%%
\[
\tilde{A}_a^0 \to \tilde{A}_a^0 + \left[ C D \left( \partial^0 
\eta_a - \tilde{g} f_{abc} \tilde{A}_b^0 \eta_c \right) \zeta 
\right]^{(\epsilon)} \, . 
\]
%%%%%%%%%%%%%%%%%%%%%%%%%%%%%

$\tilde{\cal L}^{(\epsilon)}$ in Eq. (\ref{cat}) must be invariant 
under all the symmetries of the original Lagrangian, i.e., (1) 
invariance under the spatial rotation, (2) global gauge invariance, 
(3) antighost translation invariance, (4) ghost number conservation, 
and (5) quark number conservation. Furthermore, as stated in 
(\ref{imp}), $\tilde{\Gamma}$ does not involve $\partial_0 
\bar{\eta}$. (For completeness, this is proved in Appendix D 
through formal diagrammatic analyses.) Taking all of these facts, we 
find that the most general renormalizable interaction takes the 
form: 
%%%%%%%%%%%%%%%%%%%%%%%%%%
\begin{eqnarray}
\tilde{\cal L}^{(\epsilon)} & = & 
\tilde{\cal L}_{\mbox{\scriptsize{eff}}} + \epsilon \tilde{l} 
\nonumber \\ 
&=& {\cal L}_{\psi A}^{(\epsilon)} + Z_\eta^{(\epsilon)} (\partial_i 
\bar{\eta}_a) (\partial^i \eta_a) - g d_{abc}^{(\epsilon)} 
(\partial_i \bar{\eta}_a) A_b^i \eta_c \, , 
\label{3.10d} 
\end{eqnarray}
%%%%%%%%%%%%%%%%%%%%%%%%%%%%%
where $Z_\eta^{(\epsilon)}$, $d_{abc}^{(\epsilon)}$ are unknown 
constants, and ${\cal L}_{\psi A}^{(\epsilon)}$ is the 
renormalizable term that involves only the quark and gauge fields. 

Imposition of the invariance under the transformation 
(\ref{eei1}) - (\ref{pusai1}) yields 
%%%%%%%%%%%%%%%%%%%%%%%%%%
\begin{eqnarray}
\tilde{\cal L}^{(\epsilon)} &=& \left[ - \frac{Z_{31}}{4} 
\tilde{F}_a^{ij} \tilde{F}^a_{ij} - \frac{Z_{31}'}{2} 
\tilde{F}_a^{0i} \tilde{F}_a^{0i} + Z_{21} \bar{\psi} \left( 
i \partial^j - \tilde{g} t_a A_a^j \right) \gamma_j \psi 
\right. \nonumber \\ 
&& \left. + Z_{22} \bar{\psi} \left( i \partial^0 - \tilde{g} t_a 
\tilde{A}_a^0 \right) \gamma_0 \psi - Z_m' m \bar{\psi} \psi 
+ \tilde{Z}_3 (\partial_i \bar{\eta}_a) \left\{ \partial^i 
\eta_a - \tilde{g} f_{abc} A_b^i \eta_c \right\} 
\right]^{(\epsilon)} \, . \nonumber \\ 
\label{finale} 
\end{eqnarray}
%%%%%%%%%%%%%%%%%%%%%%%%%%%%%
The tilde on $\tilde{F}_a^{ij}$ indicates that the field strength is 
to be calculated using $\tilde{g}$ for $g$, while the tilde on 
$\tilde{F}_a^{0i}$ indicates that the field strength is to be 
calculated using $\tilde{g}$ for $g$ {\em and} $\tilde{A}_a^0$ for 
$A_a^0$. All the constants appearing above are of the form, 
$Z_{31}^{(\epsilon)} = (Z_{31}^{(\epsilon)})_N = 1 + \epsilon 
(Z_{31})_N$, $(Z_{31} / D)^{(\epsilon)} = (Z_{31} / 
D)^{(\epsilon)}_N = 1 + \epsilon \sum_{N' = 0}^N (Z_{31})_{N'} D^{- 
1}_{N - N'}$ $(D_0^{- 1} = 1)$, etc. 

For obtaining further informations on the constants appearing in Eq. 
(\ref{finale}), we make following observations: 
\begin{description}
\item{(i)} Gauge-fixing independence of the locations of the 
physical poles (zeros) in the transverse-gluon- and the 
quark-propagators (two-point effective actions). \\ 
Kobes, Kunstattar, and Rebhan \cite{KKR} showed this proposition to 
hold by using a set of identities that determine the gauge 
dependence of the effective action. 
\item{(ii)} Argument in \cite{KKR} goes as it is even in 
$d$-dimensional spacetime. 
\item{(iii)} In the case of covariant gauge, the poles of the 
transverse-gluon propagator are at $p_0 = \pm p$ ($P^2 = 0$) and the 
poles of the quark propagator are at $p_0 = \pm (p^2 + 
m_{\mbox{\scriptsize{ph}}}^2)^{1 / 2}$ ($P^2 = 
m_{\mbox{\scriptsize{ph}}}^2$), where $m_{\mbox{\scriptsize{ph}}}$ 
is the physical quark mass. Thus, the dispersion relation of the 
transverse-gluon (quark) mode is Lorentz invariant, $P^2 = 0$ 
($P^2 - m^2_{\mbox{\scriptsize{ph}}} = 0$). 
\end{description}
Our $Z^{(\epsilon)}$'s are the ones that are computed in the 
minimal subtraction scheme in dimensional regularization. For taking 
the facts (i) - (iii) into account, the on-shell renormalization 
scheme (RS) should be employed. Let 
$(Z_{31}^{(\epsilon)})_{\mbox{\scriptsize{on}}}$, etc. be the 
on-shell RS counterpart of $Z_{31}^{(\epsilon)}$, etc. Then, 
(i) - (iii) and the on-shell RS counterparts of Eq. (\ref{finale}) 
tell us that $(Z_{31}^{(\epsilon)})_{\mbox{\scriptsize{on}}} = 
(Z_{31}^{(\epsilon) '})_{\mbox{\scriptsize{on}}}$ and 
$(Z_{21}^{(\epsilon)})_{\mbox{\scriptsize{on}}} = 
(Z_{22}^{(\epsilon)})_{\mbox{\scriptsize{on}}}$. Since 
$(Z_{31}^{(\epsilon)})_{\mbox{\scriptsize{on}}} - 
Z_{31}^{(\epsilon)}$, etc. are UV finite, we obtain\footnote{For 
arriving at the strict Coulomb gauge $(\alpha \to 0)$, one might 
encounter a subtle limiting procedure. The proposition (i) has 
been proved within the interpolating gauge in \cite{Z2}.} 
%%%%%%%%%%%%%%%%%%%%%%%%%%%%%%%%%%%%%%%%%%%%%%%%%%%%%%%%%%%%%%%
\begin{equation}
Z_{31}^{(\epsilon)} = Z_{31}^{(\epsilon) '} \,, \;\;\;\;\;\; 
Z_{21}^{(\epsilon)} = Z_{22}^{(\epsilon)} \;\; \left( \equiv 
Z_2^{(\epsilon)} \right) \, . 
\label{3.11d}
\end{equation}
%%%%%%%%%%%%%%%%%%%%%%%%%%%%%
%%%%%%%%%%%%%%%%%%%%%%%%%%%%%%%%%%%%%%%%%%%%%%%%%%%%%%%%%
%%%% SEC %%%%%%%%%%%%%%%%%%%%%%%%%%%%%%%%%%%%%%%%%%%%%%%%
%%%%%%%%%%%%%%%%%%%%%%%%%%%%
\section{Algebraic renormalizability and the form of 
$\Gamma^{(\infty)}$}
In this section, we show the algebraic renormalizability of the 
Coulomb-gauge QCD. On the basis of the recursive construction in the 
last section, we deduce the form of $\Gamma^{(\infty)}$. Reviving 
the suffix $N$, setting $\epsilon = 1$, and summing over $N$, we 
obtain from Eqs. (\ref{3.10d}) and (\ref{finale}) with Eq. 
(\ref{3.11d}), 
%%%%%%%%%%%%%%%%%%%%%%%%%%%%%%%%%%%%%%%%
\begin{eqnarray} 
\tilde{\ell} & \equiv & \tilde{\cal L}_{\mbox{\scriptsize{eff}}} + 
\sum_{N = 1}^\infty \tilde{l}_N \nonumber \\ 
&=& - \frac{Z_{31}}{2} \left( \partial_\mu A_{aj} \partial^\mu A_a^j 
- \partial_i A_a^j \partial_j A_a^i \right) + Z_{32} \partial_0 
A_a^i \partial_i A_a^0 - \frac{Z_{33}}{2} \partial_i A_a^0 
\partial^i A_a^0 \nonumber \\ 
&& + g Z_{11} f_{abc} (\partial_i A_j^a) A_b^i A_c^j 
+ g Z_{12} f_{abc} \partial^0 A_a^i A_b^0 A_{c i} + g Z_{13} f_{abc} 
\partial^i A^0_a A_{bi} A_c^0 \nonumber \\ 
&& - \frac{g^2}{4} Z_{41} f_{abc} f_{ade} A_{bi} A_{cj} A^i_d A^j_e 
- \frac{g^2}{2} Z_{42} f_{abc} f_{ade} A^0_b A_{ci} A^0_d A^i_e 
\nonumber \\ 
&& + \tilde{Z}_3 (\partial_i \bar{\eta}_a ) \partial^i \eta_a - g 
\tilde{Z}_1 f_{abc} (\partial_i \bar{\eta}_a) A_b^i \eta_c \nonumber 
\\ 
&& + Z_2 \bar{\psi} \left( i 
\partial\kern-0.045em\raise0.3ex\llap{/}\kern0.25em\relax - (m - 
\delta m) \right) \psi - g Z_{\psi 11} \bar{\psi} t_a A_a^j \gamma_j 
\psi - g Z_{\psi 12} \bar{\psi} t_a A_a^0 \gamma^0 \psi \, , 
\label{dog}
\end{eqnarray} 
%%%%%%%%%%%%%%%%%%%%%%%%%%%%%%%%%%%%%%%%
where, with obvious notation, $Z_{31} = 1 + \sum_{N = 
1}^\infty (Z_{31})_N$, etc., and 
%%%%%%%%%%%%%%%%%%%%%%%%%%%%%%%%%%%%%%%%
\[ 
\begin{array}{llll} 
Z_{32} = D Z_{31} / D', \;\;\; & Z_{33} = D^2 Z_{31} / D^{'2}, 
                        \;\;\; & Z_{11} = Z_{31} / D, 
                        \;\;\; & Z_{12} = Z_{31} / D', \\ 
Z_{13} = D Z_{31} / D^{'2}, \;\;\; & Z_{41} = Z_{31} / D^2, 
                        \;\;\; & Z_{42} = Z_{31} / D^{'2}, 
                        \;\;\; & \tilde{Z_1} = \tilde{Z}_3 / D, \\ 
Z_{\psi 11} = Z_2 / D, \;\;\; & Z_{\psi 12} = Z_2 / D', 
                       \;\;\; & \delta m = (1 - Z_m' / Z_2) m \, . 
\end{array} 
\] 
%%%%%%%%%%%%%%%%%%%%%%%%%%%%%%%%%%%%%%%%
Here, $D = 1 + \sum_{N = 1}^\infty D_N$, etc. From these relations 
we obtain the identities: 
%%%%%%%%%%%%%%%%%%%%%%%%%%%%%%%%%%%%%%%%
\begin{eqnarray} 
\frac{Z_{31}}{Z_{11}} &=& \frac{\tilde{Z}_3}{\tilde{Z}_1} = 
\frac{Z_{11}}{Z_{41}} = \frac{Z_{32}}{Z_{12}} = 
\frac{Z_{33}}{Z_{13}} = \frac{Z_2}{Z_{\psi 11}} \, \; (= D) \, , 
\nonumber \\ 
\frac{Z_{31}}{Z_{12}} &=& \frac{Z_{12}}{Z_{42}} = 
\frac{Z_{32}}{Z_{13}} = \frac{Z_2}{Z_{\psi 12}} \, \; (= D') \, . 
\label{STyo}
\end{eqnarray} 
%%%%%%%%%%%%%%%%%%%%%%%%%%%%%%%%%%%%%%%%q
Recalling Eq. (\ref{henkou}), we finally have, for the density of 
$\left( \Gamma_0 + \Gamma^{(\infty)} \right)_{K_n = 0}$, 
%%%%%%%%%%%%%%%%%%%%%%%%%%%%%%%%%%%%%%%%
\begin{eqnarray} 
\ell = \tilde{\ell} - \frac{Z_{31}}{2 \tilde{\alpha}} (\partial_i 
A_a^i) (\partial_j A_a^j) \, , 
\label{dogra}
\end{eqnarray} 
%%%%%%%%%%%%%%%%%%%%%%%%%%%%%%%%%%%%%%%%
where $\tilde{\alpha} = Z_{31} \alpha$. 
We now define the bare 
fields and parameters according to 
%%%%%%%%%%%%%%%%%%%%%%%%%%%%%%%%%%%%%%%%
\begin{equation} 
\begin{array}{llll} 
A_{Ba}^j = Z_{31}^{1 / 2} A_a^j \, , \; 
& A_{Ba}^0 = Z_{33}^{1 / 2} A_a^0 \, , \; 
& \eta_B = \tilde{Z}_3^{1 / 2} \eta \, , \; 
& \bar{\eta}_B = \tilde{Z}_3^{1 / 2} \bar{\eta} \, , \\ 
\psi_B = Z_2^{1 / 2} \psi \, , \; 
& \bar{\psi}_B = Z_2^{1 / 2} \bar{\psi} \, , \; 
& g_B = Z_{11} Z_{31}^{- 3 / 2} g \, , \; 
& \alpha_B = Z_{31} \alpha \, , \\ 
m_B = m - \delta m \, . & & & 
\label{w1} 
\end{array} 
\nonumber  
\end{equation} 
%%%%%%%%%%%%%%%%%%%%%%%%%%%%%%%%%%%%%%%
Then, from Eqs. (\ref{dog}) - (\ref{dogra}) we see that $\ell$ 
may be regarded \cite{IZ} as the initial Lagrangian density 
${\cal L}_{\mbox{\scriptsize{eff}}} (A_B, \eta_B, \bar{\eta}_B, 
\psi_B, \bar{\psi}_B; g_B, m_B, \alpha_B)$ written in terms of bare 
quantities. 

By construction, the density $\tilde{\ell}$ in Eq. (\ref{dog}) is 
invariant under the transformation $\chi^n \to \chi^n + \Delta 
\chi^n$ that is obtained from Eqs. (\ref{eei1}) - (\ref{pusai1}) by 
setting $\epsilon = 1$ and summing over $N$, which reads, in obvious 
notation, 
%%%%%%%%%%%%%%%%%%%%%%%%%%
\begin{eqnarray}
\Delta A_a^i & = & SA_a^i \zeta = C \left( D \partial^i \eta_a - g 
f_{abc} A^i_b \eta_c \right) \zeta = \frac{C}{\tilde{Z}_1} 
\tilde{Z}_3^{1 / 2} \left( \partial^i \eta_{Ba} - g_B f_{abc} 
A^i_{Bb} \eta_{Bc} \right) \zeta \nonumber \\ 
& \equiv & Z_a^i sA_{Ba}^i \zeta \, , \nonumber \\ 
\Delta A_a^0 &=& SA_a^0 \zeta = C \left( D' \partial_0 \eta_a - g 
f_{abc} A_b^0 \eta_c \right) \zeta = \frac{C}{\tilde{Z}_1} 
\frac{Z_{11} \tilde{Z}_3^{1 / 2}}{Z_{12}} \left( \partial_0 
\eta_{Ba} - g_B f_{abc} A_{Bb}^0 \eta_{Bc} \right) \zeta \nonumber 
\\ 
& \equiv & Z_a^0 sA_{Ba}^0 \zeta \, , \nonumber \\ 
\Delta \eta_a &=& S\eta_a \zeta = C \left( - \frac{1}{2} g f_{abc} 
\eta_b \eta_c \right) \zeta = \frac{C}{\tilde{Z}_1} Z_{31}^{1 / 2} 
\left( -  \frac{1}{2} g_B f_{abc} \eta_{Bb} \eta_{Bc} \right) \zeta 
\equiv Z_{\eta_a} s\eta_{Ba} \zeta \, , \nonumber \\ 
\Delta \psi &=& S\psi \zeta = C \left( i g \eta_a t_a \psi \right) 
\zeta = \frac{C}{\tilde{Z}_1} \left( \frac{Z_{31} \tilde{Z}_3}{Z_2} 
\right)^{1 / 2} \left( i g \eta_{Ba} t_a \psi_B \right) \zeta \equiv 
Z_\psi s\psi_B \zeta \, , 
\label{sdf} 
\end{eqnarray}
%%%%%%%%%%%%%%%%%%%%%%%%%%%%%
where use has been made of $\tilde{g} = g / D$ and Eq. (\ref{w1}). 
Then, from Eq. (\ref{enu}) with Eqs. (\ref{4.5}), (\ref{henkan}), 
and (\ref{sdf}) we have 
%%%%%%%%%%%%%%%%%%%%%%%%%%
\begin{eqnarray}
\tilde{\Gamma}_0 + \sum_{N = 1}^\infty \tilde{\Gamma}_N^{(\infty)} 
& = & \int d^4 x \left[ \tilde{\ell} + K_n S\chi^n \right] 
\label{5d} \\ 
& \equiv & \int d^4 x \left( 
\tilde{\cal L}_{\mbox{\scriptsize{eff}}} [\chi_B, \bar{\eta}_B] + 
K_{Bn} s\chi^n_B [\chi^n_B] \right) \, , \nonumber 
\end{eqnarray}
%%%%%%%%%%%%%%%%%%%%%%%%%%
where we have used $\tilde{\ell} = 
\tilde{\cal L}_{\mbox{\scriptsize{eff}}} [\chi_B, \bar{\eta}_B]$ 
(cf. after Eq. (\ref{w1})). On the right-hand side of Eq. 
(\ref{5d}), $K_a^i$ and $\partial^i \bar{\eta}_a$ appear in the 
form, 
%%%%%%%%%%%%%%%%%%%%%%%%%%%%%%%%%%%%%%%%%%%%%%%%%%%%%%%%%%%%%%%%%%%
\begin{equation} 
\tilde{Z}_3 \left[ C \tilde{Z}_1^{- 1} K_a^i + (\partial^i 
\bar{\eta}_a) \right] \left[ \partial _i \eta_a - \tilde{g} f_{abc} 
A_{bi} \eta_c \right] \, . 
\label{siki} 
\end{equation} 
%%%%%%%%%%%%%%%%%%%%%%%%%%%
Recalling here the property (\ref{imp}), we obtain the relation $C = 
\tilde{Z}_1$. Thus, \\ 
($Z_a^i, Z_a^0, Z_{\eta_a}, Z_\psi, 
Z_{\bar{\psi}}$) cannot be arbitrarily chosen but are determined 
uniquely: 
%%%%%%%%%%%%%%%%%%%%%%%%%%
\begin{eqnarray}
& Z_a^i = \tilde{Z}_3^{1 / 2} \, , \;\;\; 
& Z_a^0 = Z_{11} \tilde{Z}_3^{1 / 2} / Z_{12} \, , \nonumber \\ 
& Z_{\eta_a} = Z_{31}^{1 / 2} \, , \;\; 
&Z_\psi = Z_{\bar{\psi}} = \left( Z_{31} \tilde{Z}_3 / Z_2 
\right)^{1 / 2} \, . 
\label{tokui}
\end{eqnarray}
%%%%%%%%%%%%%%%%%%%%%%%%%%
From these relations and Eq. (\ref{w1}), we find $K_{Bn} \chi_B^n 
= \left( Z_{31} \tilde{Z}_{3} \right)^{1 / 2} K_n \chi^n$, where 
summation over $n$ is not taken $(n = 1, 2, .. , 5)$. Note that the 
factor $\left( Z_{31} \tilde{Z}_{3} \right)^{1 / 2}$ is common for 
all $n$. This fact, together with the relation $\ell = 
{\cal L}_{\mbox{\scriptsize{eff}}} (A_B, \eta_B, \bar{\eta}_B, 
\psi_B, \bar{\psi}_B; g_B, m_B, \alpha_B)$ obtained above, justifies 
\cite{IZ} {\em a posteriori} the recursive procedure in \S3. 

It should be emphasized that we have uniquely deduced the results 
(\ref{tokui}), which is in contrast with those in \cite{Z2}. 
%%%%%%%%%%%%%%%%%%%%%%%%%%%%%%%%%%%%%%%%%%%%%%%%%
%%%% SUBSUB %%%%%%%%%%%%%%%%%%%%%%%%%%%%%%%%%%%%%%%%%%%%%
%%%%%%%%%%%%%%%%%%%%%%%%%%%%%%%%%%%%%%%%%%%%%%%%%
\subsubsection*{Derivation of $\tilde{Z}_1 = Z_{12} / Z_{31}$} 
Following \cite{Z2}, we start with the Ward identity written in 
terms of the bare quantities: 
%%%%%%%%%%%%%%%%%%%%%%%%%%
\begin{equation}
\int d^3 x \left[ \frac{\delta_R \tilde{\Gamma}_B}{\delta 
\chi^n_B} \frac{\delta_L \tilde{\Gamma}_B}{\delta K_{Bn}} \right] 
(x) 
%\nonumber \\ 
%&& \mbox{\hspace*{5ex}} 
= \partial_0 \int d^3 x \left[ \eta_{Ba} 
\frac{\delta_R \tilde{\Gamma}_B}{\delta A_{Ba}^0} + K_{A_{Ba}^0} 
\frac{\delta_L \tilde{\Gamma}_B}{\delta K_{\eta_{Ba}}} \right] (x) 
\, . 
\label{40}
\end{equation}
%%%%%%%%%%%%%%%%%%%%%%%%%%%%%
According to our results on renormalization, the effective action 
$\tilde{\Gamma}_B$ is finite \cite{Z2} when expressed in terms of 
renormalized quantities, 
%%%%%%%%%%%%%%%%%%%%%%%%%%%%%%%%%%%%%%%%%%%%%%%%%%%%%%%%%%%%%
\[
\tilde{\Gamma}_B (\chi_B, \bar{\eta}_B, K_B; g_B, m_B, \alpha_B) = 
\tilde{\Gamma} (\chi, \bar{\eta}, K; g, m, \alpha) \, . 
\]
%%%%%%%%%%%%%%%%%%%%%%%%%%%%%%%%%%%%%
Through this change of quantities, Eq. (\ref{40}) turns out to 
%%%%%%%%%%%%%%%%%%%%%%%%%%
\begin{equation}
%&& 
\int d^3 x \left[ \frac{\delta_R \tilde{\Gamma}}{\delta 
\chi^n} \frac{\delta_L \tilde{\Gamma}}{\delta K_n} \right] 
(x) 
%\nonumber \\ 
%&& \mbox{\hspace*{3ex}} 
= \frac{Z_{31} \tilde{Z}_1}{Z_{12}} 
\partial_0 \int d^3 x \left[ \eta_a \frac{\delta_R 
\tilde{\Gamma}}{\delta A_a^0} + K_{A_a^0} \frac{\delta_L 
\tilde{\Gamma}}{\delta K_{\eta_a}} \right] (x) \, . 
\label{mou}
\end{equation}
%%%%%%%%%%%%%%%%%%%%%%%%%%%%%
Since the left-hand side of this equation is UV finite, $Z_{31} 
\tilde{Z}_1 / Z_{12}$ must be finite. This implies that in the 
recursive procedure described in \S3, the UV-divergent part of 
$Z_{31} \tilde{Z}_1$ is equal to that of $Z_{12}$ in each loop 
order. Then, we obtain the relation 
%%%%%%%%%%%%%%%%%%%%%%%%%%
\begin{equation}
\tilde{Z}_1 = \frac{Z_{12}}{Z_{31}} = \frac{1}{D'} \, , 
\label{hiya}
\end{equation}
%%%%%%%%%%%%%%%%%%%%%%%%%%%%%
where use has been made of Eq. (\ref{STyo}). 
%%%%%%%%%%%%%%%%%%%%%%%%%%%%%%%%%%%%%%%%%%%%%%%%%%%%%%
%% SUBSUB %%%%%%%%%%%%%%%%%%%%%%%%%%%%%%%%%%%%%%%%%%%%%%%%
%%%%%%%%%%%%%%%%%%%%%%%%%%%%%%
\subsubsection*{Strict Coulomb gauge}
In Appendices B and D, we show that, in the strict Coulomb gauge, 
$\tilde{Z}_1 = D' = 1$, which is in accord with Eq. (\ref{hiya}). 
Using this in Eq. (\ref{STyo}), we obtain 
%%%%%%%%%%%%%%%%%%%%%%%%%%%%%%%%%%%%%%%%%%%%%
\[ 
Z_{12} = Z_{42} = Z_{31} \, , \;\;\; Z_{13} = Z_{32} \, , \;\;\; 
Z_{\psi 12} =  Z_2 \, . 
\] 
%%%%%%%%%%%%%%%%%%%%%%%%%%%%%%%%%%%%%%%%%%%%%%
Then, the relation $g_B A_{Ba}^0 = g A_a^0$ (cf. after Eq. 
(\ref{w1})) and then 
%%%%%%%%%%%%%%%%%%%%%%%%%%%%%%%
\begin{equation}
g_B^2 D_B^{00} = g^2 D^{00} \, , 
\label{jyuu}
\end{equation}
%%%%%%%%%%%%%%%%%%%%%%%%%%%%%%%%%
hold, where $D^{00}$ is the time-time component of the gluon 
propagator. 

It is worth mentioning here that, in the Hamiltonian, first-order, 
formalism \cite{Z2}, the identity (\ref{jyuu}) is deduced from the 
Ward identity. In contrast, as seen above, for deriving Eq. 
(\ref{jyuu}) in the Lagrangian formalism dealt with here, use of the 
Ward identity is not necessary. As a matter of fact, $D' = 1$ and 
$\tilde{Z}_1 = 1$ plays a role. The former is derived from the 
Zinn-Justin equation in Appendix B and the latter is obtained 
in Appendix D through formal diagrammatic analysis as in the case of 
Landau gauge. Physical importance of the relation (\ref{jyuu}) is 
fully discussed in \cite{Z1,Z2}. 

As a matter of course the Ward identity contains richer informations 
than the Zinn-Justin equation, so that different relations are to be 
obtained between various (renormalized) amplitudes. 
%%%%%%%%%%%%%%%%%%%%%%%%%%%%%%%%%%%%%%%%
%%%%%%%%% SEC %%%%%%%%%%%%%%%%%%%%%%%%%%%%%%%
%%%%%%%%%%%%%%%%%%%%%%%%%%%%%%%%%%%%%%%%
\subsubsection*{One-loop wave-function renormalization constants 
and $\delta m$ in the strict Coulomb gauge}
As an illustration, we display here the results for the 
wave-function renormalization constants and $\delta m$ to 
one-loop order in the strict Coulomb gauge (see Appendix C): 
%%%%%%%%%%%%%%%%%%%%%%%%%%%%%%%%%%%%%%%%%%%%%%%%%%%%%%%%%%%
\begin{eqnarray}
Z_{31} &=& - \frac{3 g^2}{8 \pi^2} \frac{1}{d - 4} 
+ \frac{g^2}{12 \pi^2} \frac{1}{d - 4} \, , 
\label{i} \\ 
Z_{32} &=& - \frac{7 g^2}{8 \pi^2} \frac{1}{d - 4} 
+ \frac{g^2}{12 \pi^2} \frac{1}{d - 4} \, , 
\label{ro} \\ 
Z_{33} &=& - \frac{11 g^2}{8 \pi^2} \frac{1}{d - 4} 
+ \frac{g^2}{12 \pi^2} \frac{1}{d - 4} \, , 
\label{ha} \\ 
\tilde{Z}_3 & = & - \frac{g^2}{2 \pi^2} \frac{1}{d - 4} \, , 
\label{ni} \\ 
Z_2 & = & \frac{g^2}{6 \pi^2} \frac{1}{d - 4} \, , 
\label{ho} \\ 
\delta m & = & - \frac{g^2}{2 \pi^2} \frac{1}{d - 4} \, m 
\, . 
\label{he} 
\end{eqnarray}
%%%%%%%%%%%%%%%%%%%%%%%%%%%%%
The second terms on the right-hand sides of Eqs. 
(\ref{i}) - (\ref{ha}) come from the quark-loop diagram. These 
$Z$'s and $\tilde{Z}_3$ satisfy the identities (\ref{STyo}). 

We explicitly confirmed in the Lagrangian formalism, as in other 
literatures in different formalisms, the absence of energy 
divergence in one-loop order (Appendix C). 
%%%%%%%%%%%%%%%%%%%%%%%%%%%%%%%%%%%%%%%%%%%%%%%%%%%%%%%%%
%%%%%%%% SEC %%%%%%%%%%%%%%%%%%%%%%%%%%%%%%%%%%%%%%%%%%%%
%%%%%%%%%%%%%%%%%%%%%%%%%%%%%%%%%%%%%%%%%%%%%%%%%%%%%%%%%
\section{Summary and discussion} 
In this paper we have addressed the problem of renormalizability of 
the Coulomb-gauge QCD {\em within the Lagrangian, second-order, 
formalism}. Starting with the Zinn-Justin equation and following the 
procedure as in \cite{wein}, we have proved a formal or algebraic 
renormalizability. The renormalization constants for the external 
sources $K_a^i$, $K_a^0$, $K_{\eta_a}$, $K_\psi$, and 
$K_{\bar{\psi}}$, which couple, respectively, to the composite 
operators $sA_a^i$, $sA_a^0$, $s\eta_a$, $s\psi$, and 
$s\bar{\psi}$ are uniquely determined. 

We have derived the Ward identity (\ref{chu4te}) and from which we 
have obtained the identity $\tilde{Z}_3 = Z_{32} / Z_{31}$. With the 
aid of the Zinn-Justin equation, we have shown that $g A_a^0$ is 
unchanged under renormalization. In \cite{Z1,Z2}, this is derived 
using the Ward identity. In contrast, in the present Lagrangian 
formalism, use of the Zinn-Justin equation is sufficient for proving 
this. 
%%%%%%%%%%%%%%%%%%%%%%%%%%%%%%%
%%%% ACK %%%%%%%%%%%%%%%%%%%%%%%%%%%%%%%%%%
%%%%%%%%%%%%%%%%%%%%%%%%%%%%
\section*{Acknowledgments}
I wish to thank M. Inui and H. Kohyama for useful discussions. 
This work has been supported in part by the Grant-in-Aid for 
Scientific Research [(C)(2) No. 17540271] from the Ministry of 
Education, Culture, Sports, Science and Technology, Japan, 
No.(C)(2)-17540271. 
%%%%%%%%%%%%%%%%%%%%%%%%%%%%%%%%%%%%%%%
%%%%% App %%%%%%%%%%%%%%%%%%%%%%%%%%%%%%%%%%
%%%%%%%%%%%%%%%%%%%%%%%%%%%%%%%%%%%%%%%
\begin{appendix}
%%%%%%%%%%%%%%%%%%%%%%%%%%%%%%%%%%%%%%%
%%%%% App %%%%%%%%%%%%%%%%%%%%%%%%%%%%%%%%%%
%%%%%%%%%%%%%%%%%%%%%%%%%%%%%%%%%%%%%%%
\section{Ward identity and Zinn-Justin equation} 
\setcounter{equation}{0}
\setcounter{section}{1}
\def\theequation{\mbox{\Alph{section}.\arabic{equation}}}
\subsubsection*{$\bar{\eta}_a$-dependence of $\Gamma$} 
Changing the integration variable $\bar{\eta}_a' \to \bar{\eta}_a' + 
\delta \bar{\eta}_a'$ in Eq. (\ref{chu}), we obtain 
%%%%%%%%%%%%%%%%%%%%%%%%%%
\begin{eqnarray*}
0 & = & \int {\cal D} (\chi', \bar{\eta}') \delta \bar{\eta}_a' 
\left[ - \partial_j D^j_{ab} [A'] \eta_b' + \frac{\delta_R 
\Gamma}{\delta \bar{\eta}_a} \right] e^{i \int d^4 y (...)} 
\nonumber \\ 
& = & \int {\cal D} (\chi', \bar{\eta}') \delta \bar{\eta}_a' 
\left[ - \partial^j \frac{\delta_L}{i \delta K_a^j} 
+ \frac{\delta_R \Gamma}{\delta \bar{\eta}_a} \right] 
e^{i \int d^4 y (...)} \, . 
\end{eqnarray*}
%%%%%%%%%%%%%%%%%%%%%%%%%%%%%
From this equation, we get 
%%%%%%%%%%%%%%%%%%%%%%%%%%%%%%%%%%%%%%%%%%%%%%%%%%%%%%%%%%%%%%%%%%
\[ 
\partial_j \frac{\delta_L \Gamma}{\delta K_a^j} 
+ \frac{\delta_L \Gamma}{\delta \bar{\eta}_a} = 0 \, . 
\] 
%%%%%%%%%%%%%%%%%%%%%%%%%%%%%%
This relation tells us that $\Gamma$ depends on $\bar{\eta}$ only 
through $K_a^j + \partial^j \bar{\eta}_a$. 
%%%%%%%%%%%%%%%%%%%%%%%%%%%%%%%%%%%%%%%%%%%%
%%%%%%%%%%%%%%%%%%%%%%%%%%%%%%%%%%%%%%%%%%%%
%%%%%%%%%%%%%%%%%%%%%%%%%%%%%%%%%%%%%
\subsubsection*{Derivations of Ward identity and Zinn-Justin 
equation} 
We introduce infinitesimal variations \cite{Z1,Z2} 
%%%%%%%%%%%%%%%%%%%%%%%%%%
\begin{eqnarray}
\chi^{' n} (x) & \to & \chi^{' n} + f (x_0) s\chi^{' n} (x) 
\zeta \;\;\;\;\; (n = 1, ... , 5) \, , 
\label{BRS1'} \\ 
\bar{\eta}_a' (x) & \to & \bar{\eta}_a' (x) + f (x_0) 
s\bar{\eta}_a' (x) \zeta \, , 
\label{BRS2'} 
\end{eqnarray}
%%%%%%%%%%%%%%%%%%%%%%%%%%%%%
where $s\chi^{' n}$ and $s\bar{\eta}_a'$ are as in Eqs. 
(\ref{BRS2}) - (\ref{BRS4}), and $f (x_0)$ is a $x_0$-dependent 
function. When $f$ is a constant, it reduces to the BRST 
transformation. Making this change of variables in Eq. (\ref{chu}), 
we obtain 
%%%%%%%%%%%%%%%%%%%%%%%%%%
\begin{eqnarray}
&& \int {\cal D} (\chi', \bar{\eta}') \left[ \int d^4 x \left\{ - 
\left( \frac{\delta_R \Gamma}{\delta \chi^n} s\chi^{' n} + 
\frac{\delta_R \Gamma}{\delta \bar{\eta}_a} s\bar{\eta}_a' \right) 
f + \rho [ \chi' ] \partial_0 f \right\} \right] \nonumber \\ 
&& \mbox{\hspace*{5ex}} \times \exp \left[ i \int d^4 y \left( 
{\cal L}_{\mbox{\scriptsize{eff}}} (\chi', \bar{\eta}') + K_n s 
\chi^{' n} 
%\right. \right. \, \nonumber \\ 
%&& \mbox{\hspace*{12ex}} \left. \left. 
- \frac{\delta_R 
\Gamma}{\delta \chi^n} \left( \chi^{' n} - \chi^n \right) - 
\frac{\delta_R \Gamma}{\delta \bar{\eta}_a} \left( \bar{\eta}_a' - 
\bar{\eta}_a \right) \right) \right] \nonumber \\ 
&& \mbox{\hspace*{9ex}} = 0 \, , \label{chu1}
\end{eqnarray}
%%%%%%%%%%%%%%%%%%%%%%%%%%%%%
where $\rho$ is the BRST charge: 
%%%%%%%%%%%%%%%%%%%%%%%%%%
\[
\rho {[} \chi' {]} 
= - F^{' 0i}_a sA'_{ai} + i \bar{\psi}' \gamma_0 s\psi' + K_a^0 
s\eta_a' \, . 
\]
%%%%%%%%%%%%%%%%%%%%%%%%%%%%%
We rewrite the term being proportional to $\rho \partial_0 f$ as 
%%%%%%%%%%%%%%%%%%%%%%%%%%
\begin{eqnarray}
&& \int {\cal D} (\chi', \bar{\eta}') \left[ \int d^4 x \left( - 
F^{' 0i}_a sA'_{ai} + i \bar{\psi}' \gamma_0 s\psi' + K_a^0 s\eta_a' 
\right) \partial_0 f \right] e^{i \int d^4 y ( ... )} \nonumber \\ 
&& \mbox{\hspace*{6ex}} = - \int {\cal D} (\chi', \bar{\eta}') \exp 
\left[ - i \int d^4 y \frac{\delta_R \Gamma}{\delta A_a^0} 
\left( A_{a0}' - A_{a0} \right) \right] 
\nonumber \\ 
&& \mbox{\hspace*{9ex}} \times \left[ 
\int d^4 x \left( \eta_a' 
\frac{\delta}{i \delta A_{a0}'} + K_a^0 s\eta_a' \right) \partial_0 
f \right] \nonumber \\ 
&& \mbox{\hspace*{9ex}} \times \exp \left[ i \int d^4 y \left( 
{\cal L}_{\mbox{\scriptsize{eff}}} (\chi', \bar{\eta}') + K_n s 
\chi^{' n} 
\right. \right. \, \nonumber \\ 
&& \mbox{\hspace*{9ex}} \left. \left. 
- \sum_{n \neq 2} 
\frac{\delta_R \Gamma}{\delta \chi^n} \left( \chi^{' n} - \chi^n 
\right) - \frac{\delta_R \Gamma}{\delta \bar{\eta}_a} \left( 
\bar{\eta}_a' - \bar{\eta}_a \right) \right) \right] \, . 
\label{chu2}
\end{eqnarray}
%%%%%%%%%%%%%%%%%%%%%%%%%%%%%
Carrying out the partial integration, we obtain 
%%%%%%%%%%%%%%%%%%%%%%%%%%
\begin{equation}
\mbox{Eq. (\ref{chu2})} = - \int {\cal D} (\chi', \bar{\eta}') 
\left[ \int d^4 x \left( \eta_a' \frac{\delta_R \Gamma}{\delta 
A_a^0} + K_a^0 \frac{\delta_L \Gamma}{\delta K_{\eta_a}} \right) 
\partial_0 f \right] e^{i \int d^4 y (...)} \, . 
\label{chu3}
\end{equation}
%%%%%%%%%%%%%%%%%%%%%%%%%%%%%
From the definition of $\eta_a$, we can make a replacement $\eta_a' 
\to \eta_a$ . 

From here, we proceed as follows: i) Substitute Eq. (\ref{chu3}) in 
Eq. (\ref{chu1}), ii) translate to $\tilde{\Gamma}$ by Eq. 
(\ref{henkou}), and iii) integrate by 
part with respect to $x_0$ for the term that involves $\partial_0 
f$. Then, using the arbitrariness of $f (x_0)$, we finally obtain 
the Ward identity (\ref{chu4te}), 
%%%%%%%%%%%%%%%%%%%%%%%%%%
\begin{equation}
\int d^3 x \left[ \frac{\delta_R \tilde{\Gamma}}{\delta \chi^n} 
\frac{\delta_L \tilde{\Gamma}}{\delta K_n} \right] (x) 
%\nonumber \\ 
%&& \mbox{\hspace*{6ex}} 
= \partial_0 
\int d^3 x \left[ \eta_a \frac{\delta_R \tilde{\Gamma}}{\delta 
A_a^0} + K_a^0 \frac{\delta_L \tilde{\Gamma}}{\delta 
K_{\eta_a}} \right] (x) \, . 
\label{chu4}
\end{equation}
%%%%%%%%%%%%%%%%%%%%%%%%%%%%%
Integration of Eq. (\ref{chu4}) over $x_0$ yields the Zinn-Justin 
equation (\ref{ST6}). Eq. (\ref{chu4}) holds for arbitrary $\alpha$ 
(gauge parameter). 
%%%%%%%%%%%%%%%%%%%%%%%%%%%%%%%%%%%%%%%%%%%%%%%%%%
%%%%%%%%%%%%%%%%%%%%%%%%%%%%%%%%%%%%%%%%%%%%%%%%%%%%%%%%
%%%%%%%%%%%%%%%%%%%%%%%%%%%%%%%%%%%%%%%%%%%%%%%%%%
\section{Derivation of $Z_{31} / Z_{12} = D' = 1$ in the 
strict \\ 
Coulomb gauge} 
\setcounter{equation}{0}
By functionally differentiating the Zinn-Justin equation (\ref{ST6}) 
with respect to $A_a^\mu$, $A_b^\nu$, and $\eta_c$, we obtain 
\cite{dir}, after Fourier transformation, 
%%%%%%%%%%%%%%%%%%%%%%%%%%
\begin{equation}
P_\mu \Pi^{\nu \mu} (P) = \Pi^{\nu \mu} (P) \tilde{\Pi}_\mu (P) \, , 
\label{2-pt}
\end{equation}
%%%%%%%%%%%%%%%%%%%%%%%%%%%%%
which is diagonal in color space, so that the color index is 
dropped. $\Pi^{\nu \mu}$ is the two-point gluon effective action, 
from which the gauge-fixing term is dropped. $\tilde{\Pi}_\mu (P)$ 
is defined by 
%%%%%%%%%%%%%%%%%%%%%%%%%%
\begin{equation}
i g f_{abc} \langle A^\mu_b (x) \eta_c (x) \bar{\eta}_d (y) 
\rangle_{tr} \stackrel{\mbox{\scriptsize{F.T.}}}{\longrightarrow} 
\delta_{ad} \tilde{\Pi}^\mu (P) \, , 
\label{dess}
\end{equation}
%%%%%%%%%%%%%%%%%%%%%%%%%%%%%
where the suffix \lq\lq $tr$'' stands for truncation and 
\lq\lq F.T.'' means to take Fourier transformation. The FP-ghost 
self-energy part $\tilde{\Pi} (P)$ is related to $\tilde{\Pi}_\mu 
(P)$ through $p^i \tilde{\Pi}_i (P) = \tilde{\Pi} (P)$. 

Let us introduce a tensor decomposition of $\Pi^{\nu \mu}$: 
%%%%%%%%%%%%%%%%%%%%%%%%%%%%%%%%%%%%%%%%
\begin{equation} 
\Pi^{\mu \nu} (P) = g^\nu_{\; i} g^\mu_{\; j} g^{i j} A + 
p^{\underline{\nu}} p^{\underline{\mu}} B 
%\nonumber \\ 
%&& 
+  p_0 (n^\nu p^{\underline{\mu}} + 
p^{\underline{\nu}} n^\mu) C +  n^\nu n^\mu E \, , 
\label{ahh} 
\end{equation} 
%%%%%%%%%%%%%%%%%%%%%%%%%%%%%
where $n^\mu = (1, {\bf 0})$ and $p^{\underline{\mu}} = p^\mu - 
p_0 n^\mu$. Substitution of $\Pi^{\nu \mu}_B$, which is written in 
terms of bare quantities, into the \lq\lq bare counterpart'' of Eq. 
(\ref{2-pt}) yields two equations, one of which reads  
%%%%%%%%%%%%%%%%%%%%%%%%%%%%%%%%%%%%%%%%%%%%%%%%%%%%%%%%%%%%%%%%%
\begin{equation}
\left( p^2 + \tilde{\Pi}_B \right) C_B - \left( 1 - 
\frac{\tilde{\Pi}_{B0}}{p_0} \right) E_B = 0 \, . 
\label{owae}
\end{equation}
%%%%%%%%%%%%%%%%%%%%%%%%%%%%%%%%
Another equation leads to the same result as the one obtained below. 

When expressed in terms of renormalized quantities, this equation 
should become \cite{Z2} an UV-divergence free equation. Relations 
between the bare- and re\-normal\-ized-quantities are $C_B = \left( 
Z_{31} Z_{33} \right)^{- 1 / 2} C$, $E_B = Z_{33}^{- 1} E$, and 
$p^2 + \tilde{\Pi}_B = \tilde{Z}_3^{- 1} \left( p^2 + \tilde{\Pi} 
\right)$. In Appendix D, we show in the strict Coulomb gauge that 
$\tilde{\Pi}_0$ is UV finite and $\tilde{Z}_1 = 1$. Then, 
%%%%%%%%%%%%%%%%%%%%%%%%%%%%%%%%%%%%%%%%%%%%%%%%%%%%%%%%%%%%%%%%%
\[
\tilde{\Pi}_{B0} = \frac{1}{\tilde{Z}_3} \left( \tilde{Z}_3 
Z_{33}^{1 / 2} \frac{g_B}{g} \right) \tilde{\Pi}_0 = \frac{1}{D'} 
\tilde{\Pi}_0 \, , 
\]
%%%%%%%%%%%%%%%%%%%%%%%%%%%%%%%%
where $g_B$ is $g_B$ on the left-hand side of the \lq\lq bare 
counterpart'' of Eq. (\ref{dess}). Then, Eq (\ref{owae}) turns out 
to 
%%%%%%%%%%%%%%%%%%%%%%%%%%%%%%%%%%%%%%%%%%%%%%%%%%%%%%%%%%%%%%%%%
\[
\left( p^2 + \tilde{\Pi} \right) C = \tilde{Z}_1 D' \left( 1 - 
\frac{\tilde{\Pi}_0}{p_0 D'} \right) E \, . 
\]
%%%%%%%%%%%%%%%%%%%%%%%%%%%%%%%%
Since the left-hand side is UV finite, same reasoning as in \S4 
(cf. Eq. (\ref{mou})) leads to 
%%%%%%%%%%%%%%%%%%%%%%%%%%%%%%%%%%%%%%%%%%%%%%%%%%%%%%%%%%%%%%%%%
\begin{equation}
\tilde{Z}_1 = D' = 1 \, . 
\label{BB} 
\end{equation}
%%%%%%%%%%%%%%%%%%%%%%%%%%%%%%%%
These relations holds for the strict Coulomb gauge and are in accord 
with Eq. (\ref{hiya}). 
%%%%%%%%%%%%%%%%%%%%%%%%%%%%%%%%%%%%%%%%%%%%%%%%
%%%%%%%%%%%%%%%%%%%%%%%%%%%%%%%%%%%%%%%%%%%%%%%%%%%%%%%%%%
%%%%%%%%%%%%%%%%%%%%%%%%%%%%%%%%%%%%%%%%%%%%%%
\section{Derivation of Eqs. (43) - (45)} 
\setcounter{equation}{0}
The strict Coulomb gauge gluon propagator is given by Eq. 
(\ref{cir}) with $\alpha = 0$. Straightforward computation using 
Eqs. (\ref{cir}) with $\alpha = 0$, (\ref{star}), and the forms of 
3- and 4-gluon vertices yields, for the one-loop UV-divergent 
contributions to $\Pi^{\mu \nu} (Q)$ (Eq. (\ref{ahh})) (see, also, 
\cite{Z1,and1}), 
%%%%%%%%%%%%%%%%%%%%%%%%%%
\begin{eqnarray}
A &=& \frac{g^2}{120 \pi^2} \frac{1}{d - 4} \left[ 
\left( 15 Q^2 + 96 q^2 \right) 
%\right. \nonumber \\ 
%&& \left. 
- \left( 60 Q^2 + 96 q^2 \right) + 10 Q^2 \right] \, , 
\nonumber \\ 
B &=& \frac{g^2}{120 \pi^2} \frac{1}{d - 4} \left[ 93 - 48 - 10 
\right] \, , \nonumber \\ 
C &=& \frac{g^2}{24 \pi^2} \frac{1}{d - 4} \left[ 
- 3 + 24 - 2 \right] \, , \nonumber \\ 
E &=& \frac{g^2}{24 \pi^2} \frac{1}{d - 4} \left[ 
- 15 + 48 - 2 \right] q^2 \, . 
\label{kaka}
\end{eqnarray}
%%%%%%%%%%%%%%%%%%%%%%%%%
Here, each term on the right-hand side of each equation is 
the contribution from the following one-loop diagrams: 
\begin{itemize}
\item First term $\leftarrow$ The diagram that includes two 
transverse-gluon propagators. 
\item Second term $\leftarrow$ The diagram that includes one 
transverse-gluon propagator and one $A_0$ propagator {\em and} 
the tadpole diagram. 
\item Third term $\leftarrow$ The diagram with a quark loop. 
\end{itemize}

From Eqs. (\ref{kaka}) and (\ref{dog}), we extract $Z$'s, as given 
in Eqs. (\ref{i}) - (\ref{ha}) in the text. We briefly describe how 
Eq. (\ref{kaka}) is derived. The (\ref{Einf})-type energy-diverging 
integrations appear. As mentioned in \S5, it is known that, when 
all relevant contributions are added, cancellation occurs between 
them. Here, we substantiate this for the one-loop gluon two-point 
function in the Lagrangian formalism. Energy divergences arise only 
in the transverse-gluon two-point function. Explicit computation of 
the energy-divergent contributions of the type (\ref{Einf}) yields 
%%%%%%%%%%%%%%%%%%%%%%%%%%
\begin{eqnarray}
\left( \Pi_{a b}^{i j} \right)_1 & \simeq & 3 g^2 \delta_{ab} \int 
\frac{d^{\, 4} P}{(2 \pi)^4} \frac{1}{({\bf k} - {\bf p})^2} 
\left( \delta^{ij} - 
\frac{p^i p^j}{p^2} \right) \, , 
\nonumber \\ 
\left( \Pi_{a b}^{i j} \right)_2 
& = & - 3 g^2 \delta_{ab} \delta^{ij} \int \frac{d^{\, 4} P}{(2 
\pi)^4} \frac{1}{({\bf k} - {\bf p})^2} \, , 
\nonumber \\ 
\left( \Pi_{a b}^{i j} \right)_3 
& = & 6 g^2 \delta_{ab} \int \frac{d^{\, 4} P}{(2 \pi)^4} 
\frac{p^i p^j}{p^2 ({\bf k} - {\bf p})^2} \, , \nonumber \\ 
\left( \Pi_{a b}^{i j} \right)_4 & = & - 3 g^2 \delta_{ab} \int 
\frac{d^{\, 4} P}{(2 \pi)^4} \frac{p^i p^j}{p^2 ({\bf k} - 
{\bf p})^2} \, . 
\label{2d}
\end{eqnarray}
%%%%%%%%%%%%%%%%%%%%%%%%%
Here $\simeq$ indicates the energy-divergent contribution, and 
$\Pi_j$ $(j = 1, ... ,4)$ is the contribution from the following 
one-loop diagrams: 
\begin{itemize}
\item $\Pi_1$ $\leftarrow$ The diagram that includes one 
transverse-gluon propagator and one $A_0$ propagator. 
\item $\Pi_2$ $\leftarrow$ The diagram that includes one $A_0$ 
propagator (tadpole diagram). 
\item $\Pi_3$ $\leftarrow$ The diagram that includes two $A_0$ 
propagators. 
\item $\Pi_4$ $\leftarrow$ The diagram with FP-ghost loop. 
\end{itemize}
It can readily be seen that the cancellation occurs between the four 
contributions in Eq. (\ref{2d}). 

The ill-defined integrals like (\ref{ill1}) also appear. As 
mentioned in \S5, such integrals can be set equal to zero. 
%%%%%%%%%%%%%%%%%%%%%%%%%%%%%%%%%%%%%%%
%%%%% App %%%%%%%%%%%%%%%%%%%%%%%%%%%%%%%%%%
%%%%%%%%%%%%%%%%%%%%%%%%%%%%%%%%%%%%%%%%%%%%%%%%%%%%%%%%%%%%%%
\section{Diagrammatic analyses}
\setcounter{section}{0}
In this Appendix, we {\em formally} carry out some diagrammatic 
analyses. We start with following observations. Consider a diagram 
$G$ that includes FP-ghost external lines. 
\begin{description} 
\item{a)} From Eq. (\ref{star}), we see that the vertex factor for 
the external $\bar{\eta} A^j \eta$-vertex from which the outgoing 
ghost goes out is independent of the internal loop momenta. 
\item{b)} Strict Coulomb-gauge case: From Eq. (\ref{star}), the 
vertex factor for the external $\bar{\eta} (P - Q) A^j (Q) \eta 
(P)$-vertex, into which the incoming ghost enters, is proportional 
to $p_j - q_j$. $A^j (Q)$ constitutes the gluon propagator 
(\ref{cir}), the transverse part of which includes $\delta^{j i} - 
q^j q^i / q^2$. Then we have $(p_j - q_j) (\delta^{j i} - q^j q^i / 
q^2) = p_j (\delta^{j i} - q^j q^i / q^2)$, so that this vertex 
factor turns out to be independent of the internal loop momenta. 
\end{description}
%%%%%%%%%%%%%%%%%%%%%%%%%%%%%%%%%%%%%%%%%%%%%%%%%%%%%%%%%%%%%
%%%%%%%%%%%%%%%%%%%%%%%%%%%%%%%%%%%%%%%%%%%%%%%%%%%%%%%%%%%%%
%%%%%%%%%%%%%%%%%%%%%%%%%%%%%%%%%%%%%%%%%%%%%%%%%%
\subsubsection*{UV divergent contribution to the ghost 
propagator $\tilde{\Pi} (P)$}
Consider a diagram $G$ that contributes to $\tilde{\Pi} (P)$. The 
vertex factor (Eq. (\ref{star})) for the external $\bar{\eta} (P) 
A^i \eta$-vertex from which the outgoing ghost goes out is 
proportional to $p^i$. The vertex factor for another external 
$\bar{\eta} (Q) A^j \eta$-vertex, into which the incoming ghost 
enters, is proportional to $q^j$. Because of the rotation symmetry, 
we have, after loop-integration, $\tilde{\Pi} (P) \propto p^2 
{\cal F}$, where ${\cal F}$ is a dimensionless function of $p_0$, 
$p$, and $m$. Since we have assumed that the divergent part of 
$\tilde{\Pi}$ is a local function of the fields (cf. after Eq. 
(\ref{enu})), we have $\tilde{\Pi}^{(\infty)} (P) \propto p^2 / (d - 
4)^n$ $(n = 1, 2, ... )$, so that $\tilde{\cal L}^{(\epsilon)}$ in 
Eq. (\ref{cat}) does not involve the term $(\partial_0 \bar{\eta}) 
(\partial_0 \eta)$. 
%%%%%%%%%%%%%%%%%%%%%%%%%%%%%%%%%%%%%%%%%%%%%%%%%%%%%%%%%%%%%
%%%%%%%%%%%%%%%%%%%%%%%%%%%%%%%%%%%%%%%%%%%%%%%%%%%%%%%%%%%%%
%%%%%%%%%%%%%%%%%%%%%%%%%%%%%%%%%%%%%%%%%%%%%%%%%%
\subsubsection*{UV finiteness of $\bar{\eta} A^0 \eta$ 
three-point functions}
Let ${\cal F} (P, Q)$ be a $\bar{\eta} A^0 (Q) \eta (P)$ 
three-point function. From its Lorentz structure, ${\cal F}$ is of 
the form ${\cal F} = {\cal F}_1 p_0 + {\cal F}_2 q_0$. This, 
together with the above observation a), shows that the degree of UV 
divergence is $- 1$, so that ${\cal F}$ is UV finite. Then, 
$\tilde{\cal L}^{(\epsilon)}$ in Eq. (\ref{cat}) does not involve 
the term $(\partial_0 \bar{\eta}) A^0 \eta$. 
%%%%%%%%%%%%%%%%%%%%%%%%%%%%%%%%%%%%%%%%%%%%%%%%%%%%%%%%%%%%%
\subsubsection*{UV finiteness of $\tilde{\Pi}_0 (P)$ in strict 
Coulomb gauge} 
Consider a diagram $G$ that contributes to $\tilde{\Pi}_0 (P)$, Eq. 
(\ref{dess}) with $\mu = 0$. From the Lorentz structure 
$\tilde{\Pi}_0 (P) \propto p_0$. This, together with the above 
observations a) and b), shows the degree of UV divergence of the 
diagram is $-1$. 
%%%%%%%%%%%%%%%%%%%%%%%%%%%%%%%%%%%%%%%%%%%%%%%%%%%%%%%%%%%%%
\subsubsection*{$\tilde{Z}_1 = 1$ in strict Coulomb gauge} 
Consider an $\bar{\eta} A^j \eta$ three-point function in the strict 
Coulomb gauge. Above observations a) and b) tell us that the 
$\bar{\eta} A^j \eta$ three-point function is UV finite and then, in 
the minimal subtraction scheme, $\tilde{Z}_1 = 1$, as in the Landau 
gauge case. 
\end{appendix}
%%%%%%%%%%%%%%%%%%%%%%%%%%%%%%%%
%%% Ref %%%%%%%%%%%%%%%%%%%%%%%%%%%%%%%%%%%%%%%%%%%%%%%%%%%%
%%%%%%%%%%%%%%%%%%%%%%%%%%%%%%%%
 
\end{document}